\documentclass[onecolumn,showpacs,showkeys,preprintnumbers,amsmath,amssymb]{revtex4}

\usepackage{graphicx}
\usepackage{dcolumn}
\usepackage{bm}

\begin{document}


\title{Geometric phase for degenerate states of spin-1 and spin-1/2 pair}

\author{Guo-Qiang Zhu}
 \affiliation{Zhejiang Institute of Modern Physics, Zhejiang University, Hangzhou 310027, P.R. China}


\begin{abstract}
The geometric phase of a bi-particle model is discussed.  One can
drive the system to evolve by external magnetic field, thereby
controlling the geometric phase. The relationship between the
geometric phase and the structures of the initial state is obtained.
At last we extend the results to a more general case.
\end{abstract}

\maketitle
\section{Introduction}\label{section:introduction}
Geometric phase (GP) in quantum theory attracted great interest
since Berry \cite{berry} showed that the state of quantum system
acquires a purely geometric feature in addition to the usual
dynamical phase when it is varied slowly and eventually brought back
to its initial form. The general to nonadiabatic evolution extension
was formulated by Aharonov and Anandan \cite{AA}. The extension to
noncyclic evolution was done by Samuel and Bhandara \cite{samuel}.

Uhlmann \cite{uhlmann} was the first to introduce the notion of GP
for mixed state. By considering a purification and the notion of
parallelity, he furnished a definition of GP for mixed states.
Later, Sj\"{o}qvist \emph{et al.} \cite{sjoqvist} introduced a
formalism that defines the mixed state GP with the experimental
context of quantum interferometry. This definition was verified
experimentally later \cite{jfdu}. Tong \emph{et al.} gave a
kinematic approach to define GP in mixed states undergoing nounitary
evolution \cite{tong}. The kinematic approach was also used to study
the off-diagonal geometric phases for mixed state, including
nondegenerate and degenerate cases \cite{off}. The off-diagonal
mixed state geometric phase could contain interference information
when the diagonal phase is undefined.

GP of composite system has drawn much attention during recent years.
X. X. Yi \emph{et al.} studied the geometric phase of composite
mixed state \cite{mixed}.  The connection between the GP and quantum
phase transition of many-body system was studied \cite{hamma}.  GP
can be used to detect the quantum phase transition points
\cite{carollo}. The scaling behavior of GP in the vicinity of the
quantum phase transition point were also discussed \cite{shiliang}.

On the other hand, GP for some system driven by external fields is
discussed \cite{ztang,xxyi2,aberg,tan,liang,xing}. In
Ref.\cite{xxyi}, the authors calculated the geometric phase of a
two-level system driven by a quantized magnetic field subject to
phase dephasing and found that the phase reduces to the standard GP
in the weak-coupling limit.

GPs are interesting both from a fundamental point of view and for
their applications. Due to the fact that GP depend only on some
global geometric properties, geometric quantum computation is one of
the most important \cite{falci,pachos}. The geometric quantum gate
was shown robust against decoherence \cite{LMDuan} and has built-in
fault-tolerant features.

In this paper, the GP for one two-interaction-spin model which is
composed of spin-$1$ and spin-$1/2$ under unitary evolvement will be
discussed. GP is dependent upon the initial state and the
Hamiltonian. In Sec.\ref{subsection:model}, GP for different
structure of initial states is given. The effect of magnetic field
is discussed. In Sec.\ref{summary}, GP of a more general case will
be considered.

\section{Geometric phase for spin-1 and spin-$1/2$
pair}\label{subsection:model}
 We consider a simple system of spin-1
and spin-$1/2$ with anisotropic Heisenberg coupling in an uniform
magnetic field as follows
\begin{equation} \label{hamiltonian}
H=H_0+H_1=\frac J2\left( \sigma _x\cdot S_x+\sigma _y\cdot
S_y+\Delta \sigma _z\cdot S_z\right)+B(\frac{1}{2}\sigma_z+S_z),
\end{equation}
where $\sigma$'s refer to the Pauli matrices for spin-$1/2$ and $S$'
denote the spin operators for spin-$1$, $\Delta$ is the anisotropy
factor. It is obvious that the commutator $[H_0, H_1]=0$. As one
knows, in the $^6 Li$ atom, the nucleus has spin-$1$ and the
electrons have total spin-$1/2$. This can be regarded as an example.
Throughout this paper, the spin-$1/2$ states are denoted by
$|\uparrow\rangle$ and $|\downarrow\rangle$ while the spin-$1$
states are denoted by $|\Uparrow\rangle$, $|0\rangle$,
$|\Downarrow\rangle$.  Then the evolvement matrix $U(t)=\exp(-itH)$,
where the Planck constant $\hbar$ is set to one. At time $t$, the
density matrix is described by $\rho(t)=U(t)\rho(0) U(t)^{\dagger}$.

We assume at the initial time $t=0$, the magnetic field is absent.
The initial state is the ground state of $H_0$. Then the magnetic
field is imposed to drive the system to evolve. In the following, we
will study the different cases: $J>0$ and $J<0$.

\subsection{$J>0$}\label{sub_J_b}
One can easily obtain the eigenvectors and eigenvalues of the
Hamiltonian $H_0$. We assume at the initial time $t=0$, the magnetic
field $B$ is absent, i.e., $B=0$. In this case, the coupling
constant $J$ is positive, so the ground state energy is that $\Delta
J/2$ when $\Delta<-1$ and $(-\Delta-\sqrt{8+\Delta^2}) J/4$ when
$\Delta>-1$. At the point $\Delta=-1$, it is a critical point
separating two different structures of ground states. In details,
when $\Delta<-1$, there are two corresponding eigenvectors:
$|\uparrow\Uparrow\rangle$ and $|\downarrow\Downarrow\rangle$. When
$\Delta>-1$, the ground state is also twofold degenerate and the
eigenvectors are:
\begin{eqnarray}
|\Psi_1\rangle&=&\frac{1}{F_{+}}\left(|\downarrow0\rangle-\frac{\Delta+\sqrt{\Delta^2+8}}{2\sqrt{2}}|\uparrow\Downarrow\rangle\right),\\
|\Psi_2\rangle&=&\frac{1}{F_{-}}\left(|\downarrow\Uparrow\rangle+\frac{\Delta-\sqrt{\Delta^2+8}}{2\sqrt{2}}|\uparrow0\rangle\right),
\end{eqnarray}
where $F_{\pm}$ is the normalized factor.

In the case $\Delta<-1$, the two eigenvectors
$|\uparrow\Uparrow\rangle$, $|\downarrow\Downarrow\rangle$ span a
2-dimensional eigenspace. Here the initial state is assumed to be a
pure state, i.e.,
\begin{equation}
|\psi(t=0)\rangle=\cos\theta|\uparrow\Uparrow\rangle+\sin\theta
e^{i\phi}|\downarrow\Downarrow\rangle.
\end{equation}
The density matrix of the initial state
$\rho_0=|\psi(0)\rangle\langle\psi(0)|$ has $[\rho_0,H_0]=0$, so the
state will not evolve under the Hamiltonian $H_0$. GP vanishes. In
order to drive the system to evolve and  control the GP, one can
subject external magnetic field $\mathbf{B}$ to the system at the
time $t=0^+$. It is obvious that when the external magnetic field is
added, no matter how weak it is, the degeneracy of the ground state
is destroyed, due to the Zeemann split. Our interest is to study the
evolution of the initial state $|\psi(t=0)\rangle$ under the new
Hamiltonian $H=H_0+H_1$. It is obvious that $[\rho_0,H]\neq0$. After
a cyclic evolution, $\rho(T)=\rho(0)$, then one has
\begin{equation}T_1=\frac{2n\pi}{3B},\ \  n\in
\mathbf{Z}.
\end{equation}
It is the function the external magnetic field. In the following we
restrict ourselves to $n=1$ and from this on. As one knows, for a
pure state, the GP can be defined as
\begin{equation}\label{pure}
\gamma_G[U]=\gamma_t
-\gamma_d=\arg\{\langle\psi(0)|U(\tau)|\psi(0)\rangle\}+i\int_0^{\tau}\langle\psi(0)|U(t)^{\dagger}\dot{U}(t)|\psi(0)\rangle
dt.
\end{equation}
The first term on  the right side of Eq.\ref{pure} is the total
phase and the second term corresponds to the dynamical phase.  One
can obtain that
\begin{equation}\label{p1}
\gamma_G=\gamma_t-\gamma_d=2\pi\cos^2\theta.
\end{equation}

For $\Delta>-1$, the initial state is assumed to be
\begin{equation}
|\Psi(t=0)\rangle=\cos\theta|\Psi_1\rangle+\sin\theta e^{i
\phi}|\Psi_2\rangle.
\end{equation}
The period $T_2=2\pi/B$. One can obtain the GP is
\begin{eqnarray}\label{p2}
\gamma_G=\gamma_T-\gamma_d=2\pi\sin^2\theta.
\end{eqnarray}

At the point $\Delta=-1$, the universal pure state spanned by the
four eigenvectors is given by
\begin{equation}
|\Psi\rangle=\sin\theta_1\sin\theta_2\cos\theta_3|\Psi_1\rangle+\sin\theta_1\sin\theta_2\sin\theta_3e^{i\phi_1}|\Psi_2\rangle+\sin\theta_1\cos\theta_2e^{i\phi_2}|\uparrow\Uparrow\rangle+\cos\theta_1e^{i\phi_3}|\downarrow\Downarrow\rangle.
\end{equation}
One can know the period $T_3=2\pi/B$ then the GP is obtained that
\begin{equation}\label{f4}
\gamma_G=\pi(1-3\cos^2\theta_1+\sin^2\theta_1(3\cos^2\theta_2-\cos2\theta_3\sin^2\theta_2)).
\end{equation}

 We only consider some simple
cases, for example when $\theta_1=\tan^{-1}\sqrt{3}$,
$\theta_2=\tan^{-1}\sqrt{2}$ and $\theta_3=\pi/4$, which means  the
absolute value of the coefficients of all superposed states are
$1/2$, the GP $\gamma_G=\pi$.

As we know, in realistic world, the system is subject to interact
with the environment inevitably. Due to the effect of decoherence,
the state will become mixed. The off-diagonal terms approach zero.
Before the external magnetic field is imposed, one can assume that
the initial state is mixed.  When $\Delta<-1$, the density matrix is
assumed to be
\begin{equation}
\rho(0)=a|\uparrow\Uparrow\rangle\langle\uparrow\Uparrow|+(1-a)|\downarrow\Downarrow\rangle\langle\downarrow\Downarrow|.
\end{equation}
Here $0\leq a\leq 1$.  In such a space, the density matrix is
\begin{equation}
\rho(0)=\left(
         \begin{array}{cc}
           a & 0 \\
           0 & 1-a \\
         \end{array}
       \right),
\end{equation}
the unitary operator in such a subspace is
\begin{equation}
U(t)=\left(
  \begin{array}{cc}
    \exp(-i\frac{3B+\Delta J}{2}t) & 0 \\
    0 & \exp(i\frac{3B-\Delta J}{2}t) \\
  \end{array}
\right).
\end{equation}
Using the definition of GP of degenerate mixed state given in Ref.
\cite{mixed}, one can know the GP $\gamma_G=0$ (in Sec.\ref{summary}
more details will be given). In fact $\rho_0$ and the Hamiltonian
$H$ are mutual commutative, the initial state will not varied with
the time. No matter what the coefficients are, in this case the GP
still keeps zero.


 When $\Delta>-1$, the density matrix is
assumed to be
\begin{equation}
\rho(0)=b|\Psi_1\rangle\langle\Psi_1|+(1-b)
|\Psi_2\rangle\langle\Psi_2|.
\end{equation}

At the point $\Delta=-1$, the density matrix is
\begin{equation}
\rho(0)=p_1\left(|\uparrow\Uparrow\rangle\langle\uparrow\Uparrow|+p_2|\downarrow\Downarrow\rangle\langle\downarrow\Downarrow|+p_3|\Psi_1\rangle\langle\Psi_1|+(1-p_1-p_2-p_3)|\Psi_2\rangle\langle\Psi_2|\right).
\end{equation}
In the same way, the GPs of the above states are that $\gamma_G=0$.
From above calculations, for the initial mixed state of this model,
the GP remains zero independent on the values of the coefficients.
\subsection{$J<0$}
In this subsection, we will take into account the case in which
$J<0$. The Hamiltonian can be rewritten as
\begin{equation}
H=H_0+H_1=\frac{-|J|}{2}\left( \sigma _x\cdot S_x+\sigma _y\cdot
S_y+\Delta \sigma _z\cdot S_z\right)+B(\frac{1}{2}\sigma_z+S_z),
\end{equation}

Before the initial time $t=0$, the magnetic field is absent. The
critical point here is $\Delta_C=1$. When $\Delta<1$, the ground
state energy is $(-\Delta+\sqrt{8+\Delta^2})J/4$. It is still
twofold degenerate and there are two different eigenvectors:
\begin{eqnarray}
|\phi_1\rangle&=&\frac{1}{N_1}\left(\frac{\sqrt{8+\Delta^2}-\Delta}{2\sqrt{2}}|\uparrow\Downarrow\rangle+|\downarrow 0\rangle\right),\\
|\phi_2\rangle&=&\frac{1}{N_2}\left(\frac{\sqrt{8+\Delta^2}+\Delta}{2\sqrt{2}}|\uparrow0\rangle+|\downarrow\Uparrow\rangle\right),
\end{eqnarray}
where $N_{1,2}$ are the normalization factors. The initial state is
assumed to be pure,
\begin{equation}
|\Psi(0)\rangle=\cos\theta|\phi_1\rangle+\sin\theta
e^{i\phi}|\phi_2\rangle.
\end{equation}
In the same way as in Sec.\ref{sub_J_b}, one can obtain the GP as
\begin{eqnarray}
\gamma_G=\gamma_t-\gamma_d=2\pi\sin^2\theta.
\end{eqnarray}

When $\Delta>1$, the ground state energy is $J\Delta/2$, the two
corresponding eigenvectors are $|\uparrow\Uparrow\rangle$ and
$|\downarrow\Downarrow\rangle$. The initial state is assumed to be
pure then
\begin{equation}
|\psi(t=0)\rangle=\cos\theta|\uparrow\Uparrow\rangle+\sin\theta
e^{i\phi}|\downarrow\Downarrow\rangle.
\end{equation}
One can easily obtain the GP is as the same as Eq.\ref{p1}:
$\gamma_G=2\pi\cos^2\theta$.

At the critical point $\Delta=1$, the state is fourfold degenerate.
In the same way, when the initial state is pure, GP will have the
same form as Eq.\ref{f4}.

For mixed stats, repeating the discussion in
subsection.\ref{sub_J_b}, one can know the GP remains zero for the
above three cases.

\section{Discussion and Summary}\label{summary}
In this section, one can consider a general model, $H=H_0+H_1$, in
which $H_0$ and $H_1$ are arbitrary but $[H_0,H_1]=0$. Therefore,
one has $e^{-iHt}=e^{-iH_0t}e^{-iH_1 t}$. It is assumed that before
$t=0$, one has $H=H_0$, The ground-state energy is degenerate and
the corresponding states are $|\phi_i\rangle$, $i=1\ldots  n$. The
initial state is assumed to be the linear superposition of the
eigenstates, i.e., $|\psi(0)\rangle=\sum_i c_i|\phi_i(0)\rangle$,
$\sum_i |c_i|^2=1$. Then $\rho_0=|\psi(0)\rangle\langle\psi(0)|$.
Owing to the fact $[\rho_0,H_0]=0$, the initial state does not
evolve with the time. One can disturb the system by adding $H_1$ and
$[\rho_0, H_1]\neq0$. After a cyclic evolution, one can obtain the
reduced GP:
\begin{equation}
\gamma_G=\arg\langle\Psi_0|e^{-iH_1
T}|\Psi_0\rangle+\langle\Psi_0|H_1|\Psi\rangle T.
\end{equation}
The period T satisfies $[e^{-iH_1 T},\rho_0]=0$.  The evolution is
only dependent on the $H_1$ term, and the initial state is dependent
only on the $H_0$ terms.

In realistic world, the interactions with the system and the
environment always reduce the coherence of the system. The
off-diagonal elements of the density matrix of the system approaches
zero. So one can assume the initial state is mixed,
$\tilde{\rho}_0=\sum_i p_i|\phi_i(0)\rangle\langle\phi_i(0)|$, where
$\sum_i p_i=1$. The corresponding eigenenergy is $E_0$, i.e., $
H_0|\phi_i(0)\rangle=E_0|\phi_i(0)\rangle$.

The off-diagonal geometric phase factor of the degenerate mixed
state is defined as \cite{tong}
\begin{equation}
\gamma_{\rho_{j_1}\cdots\rho_{j_l}}^{(l)}=\Phi\left[Tr(\prod_{\alpha=1}^{l}U(\tau)V_{j_{\alpha}}^{\parallel}(\tau)\sqrt[l]{\rho_{j_\alpha}(0)})\right].
\end{equation}
where $\Phi[z]=z/|z|$. If $l=1$, this reduces to the diagonal
geometric phase factor and, if $l=2$, we obtain the off-diagonal
pure state geometric phase. The eigenspace is spanned by the basis
$|\phi_k\rangle$. Here we are only to study $l=1$ order GP,
\begin{eqnarray}
U(T)=e^{-i(H_0+H_1)T}=e^{-i E_0 T}e^{-i H_1 T},\\
V^{\parallel}=e^{i E_0 T}\sum_i e^{i
T\langle\phi_k|H_1|\phi_k\rangle}|\phi_k\rangle\langle\phi_k|,
\end{eqnarray}
so that one can know the GP factor is
\begin{equation}
\gamma=e^{i\gamma_G}=\Phi[\sum_k p_k\langle\phi_k|e^{-iH_1
T}|\phi_k\rangle e^{i T\langle\phi_k|H_1|\phi_k\rangle}].
\end{equation}
It is clear when $H_1$ is diagonal in basis $|\phi_k\rangle$, i.e.,
the initial density $[\tilde{\rho}_0,H_1]=0$,
$\gamma=\exp(i\gamma_G)=1$, then the GP is zero. The model discussed
in Sec.\ref{subsection:model} is the example.

In the above, the GP for degenerate ground state is discussed. As
the parameters in the Hamiltonian changes, the ground state will
have different distinct structures. We have obtained the
relationship between GP and the initial ground states and the
magnetic field, so that one can control the GP by modulating the
magnetic field. The evolution period relies on the magnetic field.

From above, GP is dependent on the initial state and the
Hamiltonian.  For the model Eq.\ref{hamiltonian}, if one wish to
detect the GP, one should try to improve the purity of the initial
state and avoid the effect of decoherence. The external magnetic
field can be helpful to control the geometric phase, which thus
might aid in finding some applications in quantum computation.

The work was supported by NSFC No. 10674117 and 10405019.

\end{document}